\documentclass[12pt,preprint]{aastex}

\newcommand{\der}[2]{\ensuremath{ \frac{d{#1}}{d{#2}} }}
\newcommand{\dern}[3]{\ensuremath{ \frac{d^{#3}{#1}}{d{#2}^{#3}} }}
\newcommand{\dpar}[2]{\ensuremath{ \frac{\partial{#1}}{\partial{#2}} }}
\newcommand{\dparn}[3]{\ensuremath{ \frac{\partial^{#3}{#1}}{\partial{#2}^{#3}} }}
\newcommand{\intall}{\int\limits_{-\infty}^{\infty}}
\newcommand{\D}{\displaystyle}

\newcommand{\bvec}[1]{{\mbox{{\boldmath$#1$}}}} 

\newcommand{\eqnref}[1]{(\ref{#1})}

\begin{document}

\title{An Improved Method for Fitting $p$-Mode Profile Asymmetries}

\author{Bradley W. Hindman}
\affil{JILA and Department of Astrophysical and Planetary Sciences,
University of Colorado, Boulder, CO~80309-0440, USA}
\email{hindman@solarz.colorado.edu}


\begin{abstract} 

In a power spectrum of the Sun's acoustic waves, the $p$ modes have distinctly
skewed frequency profiles. Furthermore, the asymmetry is observed to have the
opposite sign in power spectra made from line-of-sight velocity and continuum
intensity. The asymmetry and its reversal in sign has previously been explained
using a combination of mechanisms that involve the acoustic source. A localized
acoustic source within an acoustic cavity naturally generates asymmetric profiles
through wave interference; however, the sign of the asymmetry due to this mechanism
is identical for all observables. The reversal of the asymmetry between velocity
and intensity observations has been attributed to the visibility of the source
itself (i.e., ``correlated noise"). In this paper, I will show that asymmetry
generated by a localized source can be interpreted as either a wave interference
effect in physical space, or a mode interference effect in spectral space. I
advocate a new mode-fitting procedure based on this new interpretation, whereby
the complex phases of all the modes determine the mode asymmetries. Further, I
suggest that information about the acoustic source function is encapsulated in
the amplitude of each mode, and present a scheme by which the source function
can be obtained from measured mode amplitudes by standard helioseismic inversion
techniques.

\end{abstract}



\section{Introduction}
\label{sec:introduction}
\setcounter{equation}{0}

\begin{figure*}%
        \epsscale{1.0}%
        \plotone{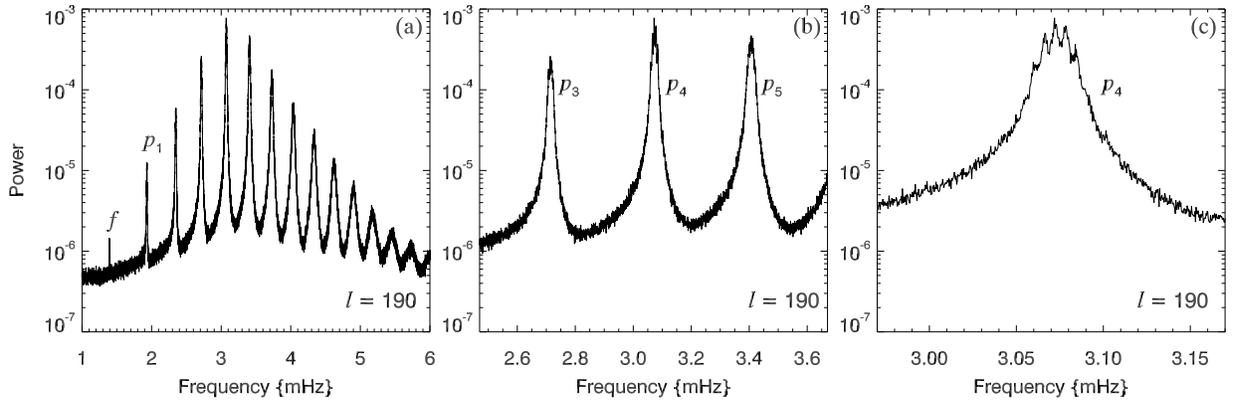}%
        \caption{\small Power spectrum as a function of frequency for a harmonic
degree of $l = 190$ and an azimuthal order of $m = 0$. The spectrum was calculated
from Dopplergram data obtained by
the GONG network. Ten years of data were separated into overlapping three month
subintervals and the individual spectra for all subintervals were averaged together.
The three panels show different frequency ranges around the $p_4$ mode.
(a) The global context showing all of the ridges that exist below the acoustic
cut-off frequency ($\nu_{\rm ac} \approx 5.2$ mHz). (b) A narrower frequency
range showing just the $p_4$ ridge and its nearest neighbors. (c) A narrow
frequency band that demonstrates that the $p_4$ ridge is a blending of the
$l = 190$, $n = 4$ mode and all of the spatial leaks from modes of nearby harmonic
degree. Differences in degree up to $\pm 2$ are clearly visible. The frequency
profile of the blended line profiles are obviously skewed or asymmetric.%
\label{fig:GONGpower_l=190}}%

\end{figure*}

\begin{figure*}%
        \epsscale{0.5}%
        \plotone{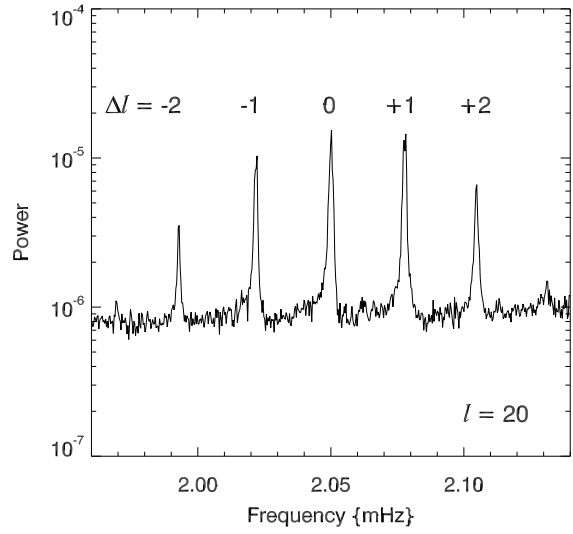}%
        \caption{\small Power spectrum as a function of frequency for waves with
$l = 20$ and $m = 0$. The same data as in Fig.~\ref{fig:GONGpower_l=190} were
used. For low values of harmonic degree (as illustrated here), each mode peak is
well-separated from the spatial leaks of nearby modes.  The leaks from modes degrees
up to $\pm 2$ are shown here. Each of the frequency profiles (for the target mode as
well as the leaks) is skewed and asymmetric.%
\label{fig:GONGpower_l=20}}%

\end{figure*}

In power spectra of solar acoustic oscillations the frequency profiles of the $p$
modes are skewed and markedly non-Lorentzian. This property was first discovered
using Ca II K observations made from the geographic South Pole \citep{Duvall:1993}
and were later confirmed \citep{Nigam:1998a} with modern spacecraft-based telescopes
using data from the Michelson Doppler Imager (MDI) onboard the Solar and Heliospheric
Observatory (SOHO). In both sets of observations, a puzzling fact was revealed: for
modes of the same frequency and harmonic degree, the asymmetry or skewness of the
line profile has opposite signs for observations made in the continuum intensity and
line-of-sight Doppler velocity. For both types of observations, the lowest-frequency
modes are the most severely skewed and the asymmetry appears to be only a weak function
of harmonic degree; however, for modes of high harmonic degree ($l \gtrsim 150$)
this statement is difficult to verify because the mode peaks are highly blended with
the spatial leaks from modes of nearby degree, forming a broad ridge of power.
Figures~\ref{fig:GONGpower_l=190} and \ref{fig:GONGpower_l=20} illustrate the asymmetry
for both high and low degree. From the figures it is evident that the frequency profile
of each individual mode is asymmetric and, at high degree, the blended ridge is also
asymmetric.

The physical mechanism for this skewness depends somewhat on whether we are considering
the asymmetry of each mode profile or the asymmetry of a blended ridge. In the former
case, the asymmetry of each mode is now thought to arise from a combination of effects
due to the nature and location of the acoustic wave source. At high harmonic degree,
the asymmetry of the $p$ mode ridge is the result of the skewness of each component
mode profile as well as an uneven distribution of power amongst the spatial leaks.
The wave source can introduce skewness in a variety of ways. The first mechanism that
I discuss here relies in detail on the location of the acoustic source. The primary
source of acoustic waves in the Sun is granular convection. Short-lived convective
events of relatively high Mach number prolifically emit sound waves.  Since granulation
is confined to a narrow band just below the photosphere, the acoustic source is highly
localized in depth. Waves generated in this layer propagate both upwards and downwards.
Those that propagate downwards are eventually refracted back upwards at the bottom of
the acoustic cavity by the sound speed gradient. When this initially downward propagating
wave returns to the excitation layer, it can destructively interfere with the wave
that was initially upward propagating. At particular frequencies, complete destructive
interference occurs and zero wave amplitude is observed above the excitation layer
(which is where the spectral lines used in actual observations are formed). The
frequencies at which the interference is complete are asymmetrically placed between
the mode ridges, thereby leading to asymmetric mode profiles \citep{Gabriel:1992,
Gabriel:1995, Roxburgh:1995, Abrams:1996, Rast:1998}.

The second mechanism due to the source arises because the convective events that generate
sound waves are themselves regions of intense temperature fluctuation and high velocity.
The sources therefore provide a contribution to the power in either intensity or velocity
observations. Since the acoustic waves are correlated with the source that generates
them, the visibility of the source can lead to power that varies significantly over
the line profile and cause skewness. This mechanism has been dubbed ``correlated noise"
and has been suggested as the reason why the asymmetry is opposite for intensity and
velocity spectra \citep{Roxburgh:1997, Nigam:1998a, Nigam:1998b}. The current understanding
is that in order to reproduce the asymmetry of both intensity and velocity power spectra
and to explain the rapid variation with frequency of the phase of the cross-spectra between
intensity and velocity \citep{Oliviero:1999}, correlated noise must provide significant
power to both types of spectra \citep[e.g.,][]{Nigam:1999, Skartlien:2000, Jefferies:2003}.

Numerical simulations of granulation have provided a counter explanation \citep{Georgobiani:2003}.
In these simulations, the velocity and intensity fluctuations have power spectra with
asymmetric mode profiles. However, if the two quantities are sampled at the same geometric
height the sign of the asymmetries are identical. Only after artificial Dopplergrams
and intensity images have been constructed using radiative transfer do the two types
of power spectra exhibit opposite asymmetries. Therefore, these simulations suggest
that details of the radiative transfer, such as the different heights at which the two
observables are formed and the nonlinear interaction between wave fluctuations and the
radiation field, are responsible for the opposite skewnesses. 

Unfortunately, many mode-fitting schemes used to measure $p$ mode frequencies assume
that the line profile is symmetric, a Lorentzian in fact. \cite{Duvall:1993} estimate
that failing to account for the asymmetry in the mode's frequency profile can lead
to mismeasurement of mode frequencies by as much as one part in $10^4$, a significant
error for the purposes of helioseismic inversions. This is probably the source of the
systematic offset ($\sim 0.1~\mu$Hz) that was observed between frequencies obtained
from intensity and velocity spectra \citep{Toutain:1997}. To account for such systematic
shifts, researchers have suggested a variety of asymmetric model profiles with which
to fit the observations \citep[e.g.,][]{Duvall:1993, Nigam:1998b, Rosenthal:1998}.
Due to its practicality, the formulation of \cite{Nigam:1998b} is probably the most
widely used, having been adapted for both the measurement of global mode frequencies
\citep{Toutain:1998, Reiter:2002, Korzennik:2005} and the fitting of ring-analysis
spectra \citep{Basu:1999, Basu:2001, Tripathy:2009}. In essence, the profile of
\cite{Nigam:1998b} is a Lorentzian times a polynomial in frequency, where the polynomial
has been calculated by performing a low order expansion about the mode frequency
$\omega_0$,

\begin{eqnarray}
	P(\omega) &=& \frac{A}{1+x^2} \, \left[s^2 + (1 + sx)^2\right] \; ,
	\label{eqn:Nigam}
\\
	x &\equiv& (\omega-\omega_0)/\gamma \; .
\end{eqnarray}

\noindent In this formula, $\gamma$ is the linewidth, $A$ is the power amplitude,
and $s$ is an asymmetry parameter.

This expression is formally valid under fairly limited conditions. For example, it has
been derived as a small argument expansion in terms of $sx$, which requires that asymmetry
is weak and more restrictively that we confine our attention to the core of the line and
avoid the wings. One clear sign of this restriction is that the profile isn't integrable,
and does not contain a finite amount of energy. Second, it ignores the presence of nearby
modes which may contribute to the power. This is a particularly onerous assumption for
high degree modes for which the line widths are sufficiently large that the mode and the
neighboring spatial leaks blend into a ridge. Figure~\ref{fig:GONGpower_l=190} illustrates
this later point quite well. Each ridge is a blending of mode peaks and the ridges are wide
enough that they have begun to overlap.

Our goal here is to generate a fitting function that takes the skewness of the
$p$-mode frequency profiles into account while maintaining its validity over a wide
range of conditions. I do so by using a modal expansion of the wavefield to represent
the power spectra using a small number of free parameters. This expansion is accomplished
in a general manner that doesn't depend on details of the atmosphere or the mechanisms
for wave damping and excitation. In the next section, Section~\ref{sec:driven}, I
discuss the driving and damping of acoustic $p$ mode oscillations in broad terms.
In Section~\ref{sec:ModeDecomp}, I derive the modal expansion of the wave field and
compute the resulting power spectrum. In Section~\ref{sec:ModeFitting}, I promulgate
a fitting profile of the spectrum that treats the asymmetry as a mode interference
phenomenon. Finally, in Section~\ref{sec:discussion}, I discuss my findings and
suggests ways in which properties of the acoustic source could be deduced through
inversion of the fitted complex mode amplitudes.


\section{Driven Acoustic Oscillations}
\label{sec:driven}
\setcounter{equation}{0}

To avoid unnecessary complexity, consider a neutrally-stable, plane-parallel
atmosphere---a reasonable approximation for modes of intermediate degree and
larger ($l \gtrsim 20$) which are trapped within the solar convection zone.
Acoustic waves in such an atmosphere obey the following wave equation:

\begin{equation}
	\left[\nabla^2 - \frac{1}{c^2}\left(\dparn{}{t}{2} + 2\gamma \dpar{}{t}
		+ \omega_c^2\right)\right]\psi(\bvec{x},t) = S(\bvec{x},t) \; ,
	\label{eqn:WaveEqnConfig}
\end{equation}

\noindent where the sound speed $c$ and acoustic cut-off frequency $\omega_c$ are
functions of height $z$. Technically, the damping rate $\gamma$ may also vary with
height, but for simplicity I will restrict my attention to a constant damping rate.
The quantity $\psi$ can represent any wave variable (with a suitable definition
for the acoustic cut-off frequency); however, in order to relate our findings to
observed power spectra made from Dopplergrams, I will treat $\psi$ as proportional
to the vertical component of the velocity. The function $S(\bvec{x},t)$ represents
the density of acoustic sources.

Since the atmosphere is invariant in the horizontal direction and in time,
Equation~\eqnref{eqn:WaveEqnConfig} can be Fourier transformed trivially to obtain
an ODE in only the height coordinate $z$,

\begin{equation}
	\left[\dern{}{z}{2} + \frac{\Omega^2(\omega)}{c^2(z)} - V(z;k)\right]\psi(z;\bvec{k},\omega)
		= S(z;\bvec{k},\omega) \; ,
	\label{eqn:WaveEqnSpec}
\end{equation}

\noindent where I adopt several definitions,

\begin{eqnarray}
	\Omega^2(\omega) &\equiv& \omega^2 + 2 i \gamma \omega \; ,
	\label{eqn:Omega2}
\\
	V(z;k) &\equiv& \frac{\omega_c^2(z)}{c^2(z)} - k^2 \; ,
	\label{eqn:Omegac2}
\end{eqnarray}

\noindent and use the following Fourier convention:

\begin{eqnarray}
	\phi(\bvec{k},\omega) = \intall\intall\intall dx \, dy \, dt \;\,
		\phi(\bvec{x},t) \, e^{-i(k_x x + k_y y - \omega t)} \; ,
\\
	\phi(\bvec{x},t) = \frac{1}{(2\pi)^3} \intall\intall\intall dk_x \, dk_y \, d\omega \;\,
		\phi(\bvec{k},\omega) \, e^{i(k_x x + k_y y - \omega t)} \; .
	\label{eqn:FConvention}
\end{eqnarray}

The frequency $\Omega$ is inherently complex and the acoustic potential $V$ is
purely real. Note, that if the complex wave frequency $\omega = \omega_r + i\omega_i$
has an imaginary part $\omega_i = -\gamma$, then the frequency $\Omega$ is purely
real with $\Omega = \sqrt{\omega_r^2 + \gamma^2}$. This allows the solution to
the damped problem to be generated by solving an equivalent undamped problem and
then shifting the resulting frequency, in both its real and imaginary components,
by an amount dependent on the damping rate, $\omega = \sqrt{\Omega^2 - \gamma^2} - i \gamma$.

\subsection{Green's Function}
\label{subsec:GFunc}

I commence by solving for the spectral Green's function $G$. For a point source located
at the height $z'$, the Green's function satisfies

\begin{equation}
	\left[\dern{}{r}{2} + \frac{\Omega^2}{c^2} - V\right] G(z,z';k,\omega) = \delta(z-z') \; .
 	\label{eqn:GreensODE}
\end{equation}

\noindent The Green's function can be found in the standard way. If I
denote $\psi_-$ as the homogeneous solution that obeys the boundary conditions
deep in the atmosphere (as $z \to -\infty$) and similarly write $\psi_+$ as the
homogeneous solution that obeys the boundary condition high in the atmosphere
(as $z \to +\infty$), the Green's function may be written in the following
form:

\begin{equation}
	G(z,z';k,\omega) = \frac{\Psi(z,z';k,\omega)}{{\cal W}(k,\omega)} \; ,
	\label{eqn:Gcontinuous}
\end{equation}

\noindent where ${\cal W}$ is the Wronskian of the two homogeneous solutions and
$\Psi$ is their product,

\begin{eqnarray}
	{\cal W}(k,\omega) &\equiv& \psi_-\der{\psi_+}{z}
		- \der{\psi_-}{z} \psi_+ \; ,
\\ \nonumber\\
	\Psi(z,z';k,\omega) &\equiv& \left\{
		\begin{array}{cl}
			\psi_-(z;k,\omega)\;\psi_+(z';k,\omega) & {\rm if} \; z<z'
		\\
			\psi_-(z';k,\omega)\;\psi_+(z;k,\omega) & {\rm if} \; z>z' \; .
		\end{array} \right.
\end{eqnarray}

Equation~\eqnref{eqn:Gcontinuous} has been used extensively in the past to explain
the asymmetry of the $p$-mode frequency profiles in the Sun's acoustic spectrum
\citep{Gabriel:1992, Roxburgh:1995, Abrams:1996, Rast:1998}. Since the driving of
acoustic waves is thought to arise from granulation, which is localized in a thin
layer just below the photosphere, I make the assumption that the height dependence
of the source function is in fact just a delta function,

\begin{equation}
	S(z;\bvec{k},\omega) = \delta(z-z') \, S_0(\bvec{k},\omega) \; .
\end{equation}

\noindent This means that the Green's function provides the correct vertical behavior
of the full wavefield.

If one further assumes that the spectral line used to observe the wavefield forms
at a height that overlies the driving layer ($z>z'$), the solution becomes

\begin{equation}
	\psi(z;\bvec{k},\omega) = S_0(\bvec{k},\omega) \frac{\psi_-(z';k,\omega)\;\psi_+(z;k,\omega)}
		{{\cal W}(k,\omega)} \; .
	\label{eqn:psi_continuous}
\end{equation}

\noindent The power spectrum of such a wavefield, observed at a height $z$ in
the atmosphere, is of course

\begin{equation}
	P(\bvec{k},\omega) = \left| \psi(z;\bvec{k},\omega) \right|^2
		= \left|S_0(\bvec{k},\omega)\right|^2 \frac{\left|\psi_-(z';k,\omega)\right|^2
			\left|\psi_+(z;k,\omega)\right|^2}
		{\left|{\cal W}(k,\omega)\right|^2} \; ,
	\label{eqn:Pcontinuous}
\end{equation}

\noindent which may be thought of as having four separate contributions: (1) the
strength of the source $\left|S_0\right|^2$, (2) a contribution from the Wronskian
$\left|{\cal W}\right|^{-2}$ that represents the modal structure of the acoustic
cavity, (3) a contribution from the driving efficiency $\left|\psi_-(z';k,\omega)\right|^2$,
and (4) a contribution from the height of observation $\left|\psi_+(z;k,\omega)\right|^2$.
The Wronskian possesses a zero at each mode frequency (which in the present formulation
will be a complex frequency because of damping). These zeros generate Lorentzian
peaks in the power spectrum at each mode frequency and by themselves these peaks
are symmetric functions of frequency. The observed asymmetry arises from the driving
efficiency which expresses how well the driver couples to any given mode. The
driving efficiency possesses zeros wherever the homogeneous solution $\psi_-$ has
a node at the driving height. This occurs whenever the downward propagating wave
generated by the source rebounds off the lower turning point of the acoustic cavity
and returns to the driving layer with the exact phase needed to destructively
interfere with the upward propagating wave also excited by the source. When this
occurs, the solution vanishes for all heights above the driving layer. The zeroes
in the driving efficiency fall in between the mode frequencies and are not distributed
symmetrically about the mode peaks; hence, they generate asymmetry in the total power
profile \citep[for complete details see][]{Rast:1998}.

\subsection{A Simple Example}
\label{subsec:SimpleExample}

To illustrate the salient features of this behavior without undue complications,
consider a simple ``guitar-string" model. Waves are trapped between two heights in
the atmosphere, $z = 0$ and $z=L$, at which the solution $\psi$ must vanish. Further,
within this cavity let the sound speed be constant and let the acoustic potential
vanish. The solutions are therefore purely sinusoidal with wavenumber $K = \Omega/c$,

\begin{eqnarray}
	\psi_-(z;\omega) &=& \sin(K z) \; ,
\\
	\psi_+(z;\omega) &=& \sin\left[K(L-z)\right] \; ,
\\
	{\cal W}(\omega) &=& K \sin(K L) \; .
\end{eqnarray}

\noindent If I assume that the driving height lies below the observation height,
and further assume that the spectral dependence of the driver is white (i.e.,
$S_0$ is a constant function of frequency), then the power spectrum of the solution
is as follows:

\begin{equation}
	P(\omega) = |S_0|^2 \frac{\sin^2(Kz') \, \sin^2\left[K(L-z)\right]}{K^2 \sin^2(KL)} \; .
\label{eqn:Pguitar_cont}
\end{equation}

The frequency dependence of this expression is illustrated in Figure~\ref{fig:Power}
with the thick black curve. The individual contributions from the Wronskian,
$|K\sin(KL)|^{-2}$ (red curve), the driving efficiency, $\sin^2(Kz')$ (blue curve)
and the observation height, $\sin^2\left[K(L-z)\right]$ (green curve), are also
indicated. The line asymmetry arises because the spacing between the zeros in
the driving efficiency is smaller than the spacing between modes (i.e., the spacing
between the zeros of the Wronskian). The most asymmetric modes (those at low
frequency) are those for which a zero in the driving efficiency almost coincides
with a mode frequency. For the model presented in the figure, the height of
observation is very close to the upper boundary of the acoustic cavity (as is
true for helioseismic observations of solar $p$ modes). This leads to a slow
frequency variation in the power due to the height of formation; therefore, this
term only provides a slow variation in the overall amplitude instead of skewness.


\section{Modal Decomposition}
\label{sec:ModeDecomp}
\setcounter{equation}{0}

While Equation~\eqnref{eqn:Pcontinuous} is useful for understanding why the Sun's
$p$-mode profiles are observed to be asymmetric functions of frequency, this
formulation for the solution $\psi$ and the resulting power spectrum $P$ does not
lend itself readily to the problem of mode fitting, whereby the mode parameters
(i.e., frequency, lifetime and amplitude) are measured from an observed spectrum.
Explicit use of Equation~\eqnref{eqn:Pcontinuous} requires prior knowledge of
the homogeneous solutions $\psi_-$ and $\psi_+$ which pre-supposes a detailed
understanding of the structure of the solar atmosphere (obtaining such an 
understanding may be the reason that one is measuring the mode frequencies in
the first place). A more natural representation of the solution would be a
modal decomposition, where the mode parameters are represented explicitly.

If one ignores the existence of a global cut-off frequency and therefore
neglects the possibility of leaky modes and continuous spectra, one may use
the orthogonality properties of the eigenfunctions of the homogeneous problem,

\begin{equation}
	\int\limits_{-\infty}^\infty dz \; \frac{\psi_n(z;k) \, \psi_m(z;k)}{c^2(z)} = \delta_{nm} \; ,
\end{equation}

\noindent to perform a standard eigenfunction expansion of the solution,
generating the following expression for the wavefield:

\begin{equation}
	\psi(z;\bvec{k},\omega) = S_0(\bvec{k},\omega) \sum_{n = 1}^\infty
		\frac{\psi_n(z';k) \, \psi_n(z;k)}{\Omega^2-\Omega_n^2(k)} \; .
	\label{eqn:psi_modal}
\end{equation}

\noindent When the atmosphere possesses a global cut-off frequency (as is true
for the Sun's $p$ modes), an eigenfunction expansion is still possible; however,
the expressions are more complicated due to the existence of continuous spectra.
To avoid obscufating the important issues with unneeded mathematics, I will consider such
possibilities separately in Appendix~\ref{app:LeakyModes}. In Equation~\eqnref{eqn:psi_modal}
the eigenvalues $\Omega_n^2$ and eigenfunctions $\psi_n$ are purely real, whereas
$\Omega^2$ is complex. This formulation is completely equivalent to that shown
previously in Equation~\eqnref{eqn:psi_continuous}; however, in this treatment
the contribution from each mode to the power is explicitly represented,

\begin{equation}
	P(\bvec{k},\omega) = \left|S_0(\bvec{k},\omega)\right|^2
		\left| \sum_{n = 1}^\infty \frac{\psi_n(z';k) \, \psi_n(z;k)}{\Omega^2-\Omega_n^2(k)} \right|^2.
	\label{eqn:Pmodal}
\end{equation}

\noindent Note that each eigenfunction $\psi_n$ is associated with two separate
complex eigenfrequencies $\omega_{\pm n}(k) = \pm \hat{\omega}_n(k) - i\gamma$,
one with a positive real part (which I will denote with the positive index $n$)
and one with a negative real part (denoted with the negative index $-n$). Therefore,
the contribution from each eigenmode can be decomposed into its two frequency
components,

\begin{eqnarray}
	\psi(z;\bvec{k},\omega) &=& S_0(\bvec{k},\omega) \sum_{n = 1}^\infty
		\frac{\psi_n(z';k) \, \psi_n(z;k)}{2\hat{\omega}_n(k)}
		\left\{\frac{1}{\omega-\omega_n} - \frac{1}{\omega-\omega_{-n}}\right\} \; ,
\\
	&=& S_0(\bvec{k},\omega) \sum_{n = -\infty}^\infty
		\frac{\psi_n(z';k) \, \psi_n(z;k)}{2\hat{\omega}_n(k)}\; 
		\frac{1-\delta_{n0}}{\omega-\omega_n(k)} \; ,
	\label{eqn:psi_freqs}
\\
	P(\bvec{k},\omega) &=& \left|S_0(\bvec{k},\omega)\right|^2
		\left| \sum_{n = -\infty}^\infty
		\frac{\psi_n(z';k) \, \psi_n(z;k)}{2\hat{\omega}_n(k)} \;
		\frac{1-\delta_{n0}}{\omega-\omega_n(k)} \right|^2.
	\label{eqn:PowerPoles}
\end{eqnarray}

\noindent The summation over $n$ now includes the negative indices, the variable
$\hat{\omega}_n \; (= -\hat{\omega}_{-n})$ is the real part of the complex mode
frequency, and the factor with the Kronecker delta $\delta_{n0}$ has been included
to remove the $n=0$ term from the summation.

The familiar representation of the power spectrum as the sum of modes, each
with a Lorentzian profile, can be recovered by assuming that all of the modes
are well-isolated from each other spectrally (the damping rate is much less
than the spacing between modes). Therefore, mode interference cross-terms can
be ignored and the power is just the incoherent sum of the power within each
mode,

\begin{equation}
	P_{\rm inc}(\bvec{k},\omega) = \left|S_0(\bvec{k},\omega)\right|^2 \sum_{n \neq 0}
			\frac{\left|\psi_n(z';k) \, \psi_n(z;k)\right|^2}{4\hat{\omega}_n^2(k)} \,
			\frac{1}{\left(\omega-\hat{\omega}_n\right)^2 + \gamma^2} \; .
	\label{eqn:Lorentzians}
\end{equation}

\noindent Of course, each Lorentzian profile in Equation~\eqnref{eqn:Lorentzians}
is symmetric about the central mode frequency $\hat{\omega}_n(k)$. The asymmetry
that is observed in $p$-mode line profiles comes from the mode interference terms
that were neglected in Equation~\eqnref{eqn:Lorentzians}.

\subsection{The Simple Example Revisited}
\label{subsec:SimpleExampleRev}

The fact that interference between modes can be construed as the source of asymmetry
in $p$-mode power profiles is aptly illustrated by the simple guitar-string model
previously explored in Section~\ref{subsec:SimpleExample}. For this simple problem, the
eigenvalues and eigenfunctions are defined by

\begin{eqnarray}
	\Omega_n &=& K_n c = \frac{n \pi c}{L} \; ,
\\
	\omega_{\pm n} &=& \pm\hat{\omega}_n - i\gamma = \pm\sqrt{\Omega_n^2-\gamma^2} - i\gamma \; , 
\\
	\psi_n(z;\omega) &=& \left(\frac{2c^2}{L}\right)^{1/2} \, \sin(K_n z) \; ,
\end{eqnarray}

\noindent and the modal expansion is just a Fourier sine series,

\begin{eqnarray}
	\psi(z;\omega) &=& \frac{c^2 S_0}{L} \sum_{n \neq 0}
		\sin(n \pi z'/L) \, \sin(n \pi z/L) \,
		\frac{\hat{\omega}_n^{-1}}{\omega - \omega_n} \; ,
\\
	P(\omega) &=& \frac{c^4 |S_0|^2}{L^2} \left| \sum_{n \neq 0}
		\sin(n \pi z'/L) \, \sin(n \pi z/L) \,
		\frac{\hat{\omega}_n^{-1}}{\omega - \omega_n} \right|^2 \; .
\label{eqn:Pguitar_mode}
\end{eqnarray}

\noindent The black curve in Figure~\ref{fig:Power} represents the power from either
representation, equations~\eqnref{eqn:Pguitar_cont} or \eqnref{eqn:Pguitar_mode}.
However, from the perspective of an eigenfunction expansion, the asymmetry in the
frequency profile of each mode arises because of mode interference. Zeroes in the
power are located at those frequencies for which complete destructive interference
has occurred, where the interference in this interpretation is in spectral space
between modes with overlapping power not in physical space between different
wave components generated by the source. Direct confirmation of this comes from
plotting the incoherent sum of the power in all the modes,

\begin{equation}
	P_{\rm inc}(\omega) = \frac{c^4 |S_0|^2}{L^2} \sum_{n \neq 0}
		\sin^2(n \pi z'/L) \, \sin^2(n \pi z/L) \,
		\frac{\hat{\omega}_n^{-2}}{(\omega-\hat{\omega}_n)^2 + \gamma^2} \; .
\end{equation}

\noindent In Figure~\ref{fig:Power} this sum is shown as the orange curve and
it can clearly be seen that the frequency profiles are symmetric, albeit modified
by a slowly varying background (which is formally composed of the wings of many
modes).


\section{Mode Fitting}
\label{sec:ModeFitting}
\setcounter{equation}{0}

In a traditional procedure, to measure properties of a $p$ mode such as its frequency,
amplitude and lifetime, the power within a narrow band of frequencies centered on the
target mode is fit with a model function. The range of frequency is selected with the
goal of mitigating the contamination from other modes. However, since spatial leakage
produces spectral peaks that cannot be fully isolated from the target mode (see
Figure~\ref{fig:GONGpower_l=190}), the fitting function is usually comprised of an
incoherent sum of power peaks based on the properties of Equation~\eqnref{eqn:Lorentzians}.
For the $j$th mode (where $j$ indicates a particular combination of harmonic degree,
azimuthal order and mode order) the fitting function might have the form, 

\begin{equation}
	P_{\rm model}(\omega) = B(\omega) + {\cal P}_j(\omega) + \sum_{j'} C_{jj'} \, {\cal P}_{j'}(\omega) \;
\end{equation}

\noindent where each profile ${\cal P}_j(\omega)$ is a Lorentzian function with three
parameters,

\begin{equation}
	{\cal P}_j(\omega) \equiv \frac{A_j}{(\omega-\hat{\omega}_j)^2 + \Gamma_j^2} \; .
\end{equation}

\noindent In these expressions, $B$ is a background function that represents incoherent
solar noise, and the three parameters that describe the Lorentzian are its mode frequency
$\hat{\omega}_j$, power amplitude $A_j$ and line width $\Gamma_j$. The leakage is described
by the leakage matrix $C_{jj'}$ and the summation is performed over all modes that have a
frequency within the fitting window and that have significant leakage. All of the free
parameters in this formulation are treated as real quantities.

Modern fitting procedures \citep[e.g.,][]{Korzennik:2005} often attempt to account for
the observed skewness in the mode profiles by using the asymmetric profile of \cite{Nigam:1998b},
Equation~\eqnref{eqn:Nigam}. Using the same notation presented here, Equation~\eqnref{eqn:Nigam}
becomes

\begin{equation}
	{\cal P}_j(\omega) = \frac{A_j}{(\omega-\hat{\omega}_j)^2 + \Gamma_j^2}
		\left\{s_j^2 + \left[1+s_j\left(\frac{\omega-\hat{\omega}_j}{\Gamma_j}\right)\right]^2\right\} \; ,
\label{eqn:NigamProfile}
\end{equation}

\noindent which adds one additional parameter, the asymmetry parameter $s_j$, to each
fitted peak.

As mentioned previously, this commonly used skewed Lorentzian function
is only a valid represention within the core of the line and it does not include the
effects of mode interference (in spectral space). Further, the energy contained by any
given mode (and hence the amplitude parameter $A_j$) is poorly defined since each mode's
power profile lacks a finite integral. A globally valid expression that directly accounts
for interference effects is suggested by the functional form of Equation~\eqnref{eqn:Pmodal},

\begin{equation}
	P_{\rm model}(\omega) = B(\omega) + \left| \frac{a_j}{\omega-\hat{\omega}_j + i\Gamma_j} +
		\sum_{j'} c_{jj'} \, \frac{a_{j'}}{\omega-\hat{\omega}_{j'} + i\Gamma_{j'}}\right|^2 \; ,
\label{eqn:FitFunc}
\end{equation}

\noindent where the parameters all have the same meaning except that the mode amplitude
$a_j$ and the leakage matrix $c_{jj'}$ are now complex quantities. Unlike classical fitting
functions, this model function properly captures asymmetric mode profiles in both the wing
and core because the effects of mode interference are present through the retention of the
cross-terms between modes. The cost of accounting for the asymmetry is the need to fit one
additional parameter per mode, the complex phase of the amplitude $a_j = |a_j| e^{i\phi_j}$.


\section{Discussion}
\label{sec:discussion}
\setcounter{equation}{0}

I have suggested an alternate formula---Equation~\eqnref{eqn:FitFunc}---for
fitting $p$-mode frequency profiles that is based on an eigenfunction expansion
of the Green's function. This formulation takes into account the skewness of the
line profiles that is inherently produced by a localized acoustic source by including
the effects of mode interference in spectral space. In addition, the fitting formula
has the salutory features that it is valid for all frequencies across the mode
profile (not just in the core of a line) and each mode contains a finite amount
of energy.

The suggested scheme can be applied to the fitting of global modes of relatively
low harmonic degree ($l \lesssim 200$) directly as envisioned. The asymmetry in a
mode's profile is primarily the result of interference with nearby spatial 
leaks, which are most likely from the same mode order or ridge (except at very low
harmonic degree $l \lesssim 10$). On the other hand, the proposed fitting model
is not immediately relevant to the fitting of ring-analysis spectra. The asymmetry
observed in the $p$-mode ridges at large harmonic degree ($l \gtrsim 200$) are not
produced by the same mechanisms that generate the asymmetry for global modes. Instead,
the asymmetry is largely a result of the spatial window function (the ring-analysis
version of the blending of a mode and the nearby leaks).

Ring-analysis spectra
are manufactured by Fourier transforming the wavefield observed on a small patch
on the surface of the Sun. The small spatial domain translates into a broad spread
of power in horizontal wavenumber. This widens the ridges and produces an effective
line width that is much larger than one would expect based just on the mode lifetime.
Furthermore, the resulting line profile depends on the window function applied
during the Fourier analysis, the slope of the $p$-mode ridge in the dispersion
diagram ($d\omega_n/dk$), and the variation in power along the ridge. Unlike
in the global mode analyses where the effects of leakage have been considered
in detail for over a decade, leakage has not been treated self-consistently
in any ring-analysis studies. While the fitting formula presented here could be
used to model the resulting ring-analysis spectra quite well with the asymmetry
being produced by the interference between the ridges themselves, the underlying
physical mechanism would be wrong. I leave it to a subsequent paper to include the
effects of the leakage function on mode profiles.

\subsection{Measuring the Acoustic Source Function}
\label{subsec:SourceFunc}

One novel property of the fitting function given by Equation~\eqnref{eqn:FitFunc}
is that the effect of the source is directly represented by the complex mode amplitudes.
Therefore, the additional parameters have a physical significance that can be used to
ferret out information about acoustic sources and the convection that engenders them.
This can be seen most easily if I relax the assumption that the source is proportional
to a delta function in height. If instead the source is distributed, as usual the
wavefield is obtained through a convolution of the Green's function with the source,

\begin{eqnarray}
	\psi(z;\bvec{k},\omega) &=& \sum_{n \neq 0} \frac{a_n(\bvec{k}, \omega)}{\omega-\hat{\omega}_n(k)} \; ,
\\
		&=& \sum_{n \neq 0} \intall dz' \; \frac{\psi_n(z';k) \, \psi_n(z;k)}{2\hat{\omega}_n(k)}\; 
		\frac{S(z';\bvec{k},\omega)}{\omega-\omega_n(k)} \; .
\end{eqnarray}

\noindent Assuming that the source function is isotropic with a frequency variation
that is slow compared to the damping rate, the complex mode amplitude is therefore
given by an integral over the source,

\begin{eqnarray}
	a_n(k) &=& \intall dz' \; {\cal K}_n(z';k) \, S(z';k,\omega_n) \; ,
	\label{eqn:Aintegrals}
\\
	{\cal K}_n(z';k) &\equiv& \frac{\psi_n(z;k) \, \psi_n(z';k)}{2\hat{\omega}_n(k)} \; ,
\end{eqnarray}

\noindent where the set of ${\cal K}_n$ are sensitivity kernels. Given a set of
measured amplitudes, the coupled integral equations~\eqnref{eqn:Aintegrals} could
be inverted using standard helioseismic inversion procedures to obtain the source
as a function of height in the atmosphere.

\subsection{Effects of Correlated Noise}
\label{subsec:CorrNoise}

The theory and procedures that have been developed in this paper have so far ignored
the existence of correlated noise. Previous studies have shown that correlated noise
is a necessary ingredient to explain the reversal of asymmetry between Doppler-velocity
spectra and continuum-intensity spectra \citep{Roxburgh:1997, Nigam:1998a, Nigam:1998b},
as well as the rapid phase variation in velocity--intensity phase difference spectra
\citep{Skartlien:2000}. The modal expansion presented here can easily be adapted to
include the effects of correlated noise \citep[e.g.,][]{Skartlien:2000}; one simply
adds a complex correlated noise component, $\psi_{\rm corr}$, to the wavefield before
computing the power,

\begin{equation}
	P_{\rm model}(\omega) = B(\omega) + \left| \frac{a_j}{\omega-\hat{\omega}_j + i\Gamma_j} +
		\sum_{j'} c_{jj'} \, \frac{a_{j'}}{\omega-\hat{\omega}_{j'} + i\Gamma_{j'}}
		+ \psi_{\rm corr} \right|^2 \; ,
\label{eqn:PowerCorrNoise}
\end{equation}

\noindent Assuming that the correlated noise is due to the visibility of the source
itself, one would expect that the phase and amplitude of the correlated noise $\psi_{\rm corr}$
would be a slowly varying function of frequency. Therefore, if one is fitting the
power in a narrow frequency band, it can safely be assumed that the noise is a complex
constant. Thus, in any fit to data the correlated noise adds only two additional
parameters that must be fit.  If one were fitting a broad range of frequencies,
it might be necessary to parameterize the frequency dependence of the correlated
noise. Since, the frequency variation should be rather weak, almost any smooth
function with a only a few free parameters (polynomial, power law, gaussian, etc.)
would probably work.


\acknowledgments

I thank Mark Rast for his help in understanding the effects of correlated noise.
I thank Irene Gonz\'alez-Hern\'andez for providing the GONG data used to generate
Figures 1 and 2. This work was supported by NASA through grants NNX08AJ08G, NNX08AQ28G,
and NNX09AB04G.

\appendix

\section{Leaky Modes}
\label{app:LeakyModes}
\setcounter{equation}{0}

The expansion of the wavefield into a discrete set of eigenmodes is appropriate
only if the system lacks a continuous spectrum. Such spectra appear as branch cuts
in the complex frequency plane. Here, I will consider two possible origins of such
branch cuts. First, temporal damping of the form appearing in Equation~\eqnref{eqn:WaveEqnConfig}
admits the possibility of overdamped solutions (OD). Such solutions manifest as a
branch cut along the imaginary frequency axis. The existence of this branch cut is
evident when one considers the frequency expression $\Omega = \sqrt{\omega^2 +2i\gamma\omega}$.
The square root function possesses branch points wherever its argument vanishes.
Therefore, there are two branch points, one at $\omega = 0$ and the other at
$\omega = -2i\gamma$, with a corresponding branch cut connecting them along the
imaginary axis. Frequencies along this branch cut correspond to overdamped waves
whose temporal dependence is pure decay without oscillation.

The second type of branch cut appears when the atmosphere possesses a global cut-off
frequency, as the Sun does. If the product of the acoustic potential and the square
of the sound speed remains bounded ($c^2 V < \alpha^2$) as the height
becomes large ($z \to \infty$), then waves of sufficiently large frequency
($\Omega > \alpha$) are not trapped and leaky waves (LW) which radiate to infinity
become possible. In the Sun, the condition for leaky waves is
$\omega/2\pi \gtrsim 5.2$ mHz. There are two branch points associated with such waves,
each located in the lower half of the complex frequency plane where $\Omega^2 = \alpha^2$
or $\omega = \pm \sqrt{\alpha^2 - \gamma^2} - i\gamma$. I choose to connect these
branch points seperately to infinity as shown in Figure~\ref{fig:omegaSpace}, and
therefore the branch cut is bipartite. Frequencies along this branch cut correspond
to propagating waves which radiate upwards through the photosphere, forming the
high-frequency, pseudo-mode portion of the Sun's acoustic spectrum.

When leaky waves are allowed by the boundary conditions, the system will not possess
an infinite number of discrete eigenmodes. Instead, for a particular horizontal
wavenumber $k$, the acoustic cavity will only have a finite number of discrete,
trapped modes and a continuous spectrum of leaky modes. As before the trapped modes
correspond to a zero in the Wronskian ${\cal W}$, or equivalently to a simple pole
in the Green's function. The leaky modes on the other hands correspond to the solutions
along a branch cut. In an eigenfunction decomposition both types of waves are needed,
as the discrete trapped waves on their own do not form a complete set. For the problem
at hand, these separate contributions can be calculated by Fourier transforms and
contour integration in the complex frequency plane, which I perform in this appendix.

I begin by taking the inverse Fourier transform of the Green's function as expressed
in Equation~\eqnref{eqn:Gcontinuous},

\begin{eqnarray}
	G(z,z';k,t) &=& \frac{1}{2\pi}\intall d\omega \; G(z,z';k,\omega) \,
		e^{-i\omega t} \;,
\nonumber
\\
	&=& \frac{1}{2\pi}\int\limits_C d\omega \;
		\frac{\Psi(r,r';k,\omega)}{{\cal W}(k,\omega)} \, e^{-i\omega t} \;.	
\end{eqnarray}

\noindent The integral is a contour integral through the complex frequency plane
along the contour $C$ which lies initially along the real frequency axis---see
Figure~\ref{fig:omegaSpace}(a). Since all waves are damped, all of the poles
and branch cuts appear in the lower half of the complex frequency plane. Therefore,
for $t<0$ one may freely deform the contour upward to infinity and the integrand
will vanish exponentially for large imaginary frequency, leading to the conclusion
$G = 0$. However, for $t>0$, in order for the integrand to vanish once again as
the countour is deformed to infinity, the deformation must be downwards into
the lower half plane. Since the lower half plane contains the poles and branch
cuts, the contour must be deformed around them and the integral acquires
contributions from the integration around each singularity. Using the residue
theorem to evaluate the integration around each pole, one obtains the contribution
from each trapped mode,

\begin{eqnarray}
	G(z,z';k,t) &=& -i \sum_{n=-N}^{N} \frac{\psi_n(z';k) \, \psi_n(z,k)}
		{{\cal W}^{\D\prime}(k,\omega_n)} \;
		(1-\delta_{n0}) \, H(t) \, e^{-i\omega_n t}
\nonumber
\\
	& & \qquad + \frac{1}{2\pi} \int\limits_{C_{\rm BC}} d\tilde{\omega} \;
		\frac{\Psi(z,z';\tilde{\omega})}{{\cal W}(k,\tilde{\omega})} \; H(t) \,
		e^{-i\tilde{\omega} t} \; ,	
\end{eqnarray}

\noindent where $H(t)$ is the Heaviside step function and the prime on the Wronskian
denotes differentiation with respect to the frequency argument,

\begin{equation}
	{\cal W}^{\D\prime} \equiv \frac{d{\cal W}}{d\omega} \; .
\end{equation}

\noindent Further, for future convenience, I have replaced the dummy variable of
integration with $\tilde{\omega}$ and the integration contour is disjoint and surrounds
all of the branch cuts $C_{\rm BC} = C_{\rm OD} + C_{\rm LW}$. The summation over
the index $n$ is over a finite number ($2N$) of poles, which includes two poles
per mode, one with a positive real part of the frequency and the other with a negative
real part, $\omega_{\pm n} = \pm\hat{\omega}_n - i\gamma$. For clarity of notation,
there is no pole associated with the index $n=0$, and I have therefore included the
factor with the Kronecker delta to exclude the $n=0$ term from the summation.

Now that I have the Green's function in the form of an eigenfunction expansion,
the result can be compared with Equation~\eqnref{eqn:psi_modal} by taking the
temporal Fourier transform and including the effects of the source,

\begin{eqnarray}
	G(z,z';k,\omega) &=& \sum_{n=-N}^N \frac{\psi_n(z';k) \, \psi_n(z,k)}
		{{\cal W}^{\D\prime}(k,\omega_n)} \;
		\frac{1-\delta_{n0}}{\omega-\omega_n(k)}
\nonumber
\\
	& & \qquad + \frac{1}{2\pi} \int\limits_{C_{\rm BC}} d\tilde{\omega} \;
		\frac{\Psi(z,z';\tilde{\omega})}{{\cal W}(\tilde{\omega})} \; \frac{1}{\omega-\tilde{\omega}} \;,
\\
	\psi(z,z';\bvec{k},\omega) &=& S_0(\bvec{k},\omega)
		\sum_{n=-N}^N \frac{\psi_n(z';k) \, \psi_n(z,k)}{{\cal W}^{\D\prime}(k,\omega_n(k))} \;
		\frac{1-\delta_{n0}}{\omega-\omega_n}
\nonumber
\\
	& & \qquad + \frac{S_0(\bvec{k},\omega)}{2\pi} \int\limits_{C_{\rm BC}} d\tilde{\omega} \;
		\frac{\Psi(z,z';\tilde{\omega})}{{\cal W}(\tilde{\omega})} \; \frac{1}{\omega-\tilde{\omega}} \; .
	\label{eqn:psi_expansion}	
\end{eqnarray}

\noindent In the equations above I have used the following Fourier transform
identity,

\begin{equation}
	\intall dt \; H(t) \, e^{i(\omega-\omega_n) t} = \frac{i}{\omega-\omega_n} \; .
\end{equation}

One can easily verify that modulo the normalization of the eigenfunctions,
Equation~\eqnref{eqn:psi_expansion} is equivalent to the modal expansion appearing
in Section \ref{sec:ModeDecomp}, Equation~\eqnref{eqn:psi_freqs}, except for the
appearance of the integral around the branch cuts (continuous spectra of leaky
waves) and a summation over only a finite number of trapped modes instead of
a countable infinity.

Finally, note that due to the symmetries of the Wronskian ${\cal W}$ and the
function $\Psi$, the integral around the branch cut associated with overdamped
waves can be shown by direct evaluation to vanish. This is true for any problem
in which the overdamped waves are confined to a cavity of finite extent. In
truly infinite domains the precise cancellation around the contour does not
occur and the overdamped solutions may play a role. Therefore, the contour
integral need only include the spectrum of leaky waves, $C_{\rm BC} = C_{\rm LW}$.

\subsection{Leaky Square-Well Potential}
\label{subsec:LeakyWellPot}

To illustrate the mathematics developed previously in this appendix, I extend the
rudimentary guitar-string model presented in the main body of this paper. Instead
of requiring that all waves reflect from the cavity boundaries, I will examine a
square-well potential that permits high-frequency waves to escape out the top of
the cavity. Consider a one-dimensional acoustic cavity of length $L$ with constant
sound speed and a global cut-off frequency. The acoustic potential $V(z)$ is
identically zero within the cavity itself, and takes on a constant, nonzero value
above the top of the cavity, $z>L$,

\begin{equation}
	V(z) = \left\{
			\begin{array}{cl}
				0 & {\rm if} \; z<L
			\\
				\alpha^2/c^2 & {\rm if} \; z>L \; .
			\end{array} \right.
\end{equation}

\noindent At the lower boundary of the cavity, $z = 0$, the solution must vanish,
and at the top of the cavity the solution must match smoothly onto either an evanescent
solution (for $\Omega < \alpha$) or an upward travelling wave (for $\Omega > \alpha$).
If I assume that the driving layer lies within the cavity itself ($z'<L$), the
homogeneous solutions $\psi_-$ and $\psi_+$ which obey the proper boundary conditions
are as follows:

\begin{eqnarray}
	\psi_-(z;\omega) &=& \sin(Kz)
\\ \nonumber \\
	\psi_+(z;\omega) &=& \left\{
	\begin{array}{ll}
		\cos\left[K(L-z)\right] + \kappa K^{-1}\sin\left[K(L-z)\right] & {\rm if} \; z<L
	\\
		\exp\left[-\kappa(z-L)\right] & {\rm if} \; z>L \; ,
	\end{array}
		\right.
\end{eqnarray}

\noindent where the constants $K$ and $\kappa$ are, respectively, the wavenumber
within the cavity and the evanescence rate for the region above the cavity,

\begin{eqnarray}
	K(\omega) &=& \frac{\Omega(\omega)}{c} \; ,
\\
	\kappa(\omega) &=& \frac{\sqrt{\alpha^2-\Omega^2(\omega)}}{c} \; .
	\label{eqn:kappa}
\end{eqnarray} 

\noindent Note, if $\Omega^2 > \alpha^2$ the evanescence rate $\kappa$ becomes
purely imaginary. For frequencies lying on the top of the branch cut, I choose the
sign of the square root in Equation~\eqnref{eqn:kappa} such that the imaginary part
of $\kappa$ is positive and the solution in the regime above the cavity, $z>L$, corresponds
to an outward propagating wave and a radiation boundary condition is satisfied.
Remember, the Fourier convention, Equation~\eqnref{eqn:FConvention}, is consistent
with solutions with the time dependence $\sim e^{-i\omega t}$.

The Wronskian of these two homogeneous solutions is of course sinusoidal,

\begin{equation}
	{\cal W}(\omega) = -\left[K\cos(KL) + \kappa\sin(KL)\right] \; .
\end{equation}

\noindent If I assume that the observations are made above the cavity, $z>L$, as
occurs for helioseismic observations, the Green's function has the solution

\begin{equation}
	G(z,z';\omega) = -\frac{\sin(Kz') \, \exp\left[-\kappa(z-L)\right]}{K\cos(KL) + \kappa(KL)} \; .
\end{equation}

\noindent Figure~\ref{fig:LeakyWell}(a) shows the power spectrum and its components
that result for this model when one assumes that the source is white (i.e., $S_0$ is 
constant),

\begin{equation}
	P(\omega) = |S_0|^2 \frac{\sin^2(Kz') \, \exp\left[-2\kappa(z-L)\right]}{{\cal W}^2(\omega)} \; .
\label{eqn:PLeaky_cont}
\end{equation}

\noindent The effect of the height of observation, $\exp\left[-2\kappa(z-L)\right]$
(green curve), is a monotonically increasing function of frequency. The contribution
from the Wronskian, $|{\cal W}|^{-2}$ (the red curve), behaves differently above and
below the global cut-off frequency. Below the cut-off, a Lorentzian spike occurs at
each mode frequency; above the cut-off, the Wronskian term oscillates and decays.
The blue curve shows the driving efficiency, $\sin^2(Kz')$. It too is oscillatory,
but with more rapid frequency variation than the Wronskian because $z' < L$. The zeros
in the driving efficiency lead to the asymmetry in the total power profile (black curve).

In the alternate representation of a modal expansion, the Green's function and the
power spectrum take on the form,

\begin{eqnarray}
	G(z,z';\omega) &=& -\sum_{n=-N}^N \frac{K_n\kappa_n^2 c^4}{1+\kappa_n L} \;
		\frac{\sin(K_n z') \, \exp\left[-\kappa_n(z-L)\right]}{\hat{\omega}_n \, \cos(K_n L)} \;
		\frac{1-\delta_{n0}}{\omega-\omega_n}
\nonumber
\\
	& & \qquad - \frac{1}{2\pi} \int\limits_{C_{\rm BC}} d\tilde{\omega} \;
		\frac{\sin(\tilde{K}z') \, \exp\left[\tilde{\kappa}(z-L)\right]}
		{\tilde{K}\cos(\tilde{K}L) + \tilde{\kappa}\sin(\tilde{K}L)} \;
		\frac{1}{\omega-\tilde{\omega}} \;,
\\
\nonumber
\\
	P(\omega) &=& \left|S_0\right|^2 \; \left|G(z,z';\omega)\right|^2 \; ,
\label{eqn:PLeaky_modes}
\end{eqnarray}

\noindent where the eigenfrequencies $\omega_n$ (or equivalently $\Omega_n$), wavenumbers
$K_n$ and evanescence rates $\kappa_n$ are defined by the roots of the Wronskian, which
provides a transcendental equation,

\begin{eqnarray}
	{\cal W}(\omega_n) &\equiv& -\left[K_n\cos(K_n L) + \kappa_n \sin(K_n L)\right] = 0 \; ,
\\
	K_n &\equiv& \frac{\Omega_n}{c} \; ,
\\
	\kappa_n &\equiv& \frac{\sqrt{\alpha^2 - K_n^2 c^2}}{c} \; .
\end{eqnarray}

\noindent  The wavenumber and evanescent length with the tilde, $\tilde{K}$ and
$\tilde{\kappa}$, are those corresponding to the integration frequency $\tilde{\omega}$,
i.e., $\tilde{K}=K(\tilde{\omega})$ and $\tilde{\kappa}=\kappa(\tilde{\omega})$.
Finally, the derivative of the Wronskian, evaluated at the mode frequencies, was
used in these expressions,

\begin{equation}
	{\cal W}^{\D\prime}(\omega_n) = - \frac{\hat{\omega}_n}{c^4} \,
		\frac{1+\kappa_n L}{K_n \kappa_n} \cos(K_n L) \; .
\end{equation}

Figure~\ref{fig:LeakyWell}(b) shows the resultant power spectrum (thick black curve) as
well as the power associated just with the modes $P_{\rm modes}$ (violet curve) 
and with the continuous spectrum $P_{\rm LW}$ (aqua curve),

\begin{eqnarray}
	P_{\rm modes} &=& \left|\sum_{n=-N}^N \frac{K_n\kappa_n^2 c^4}{1+\kappa_n L} \;
		\frac{\sin(K_n z') \, \exp\left[-\kappa_n(z-L)\right]}{\hat{\omega}_n \, \cos(K_n L)} \;
		\frac{1-\delta_{n0}}{\omega-\omega_n} \right|^2 \; ,
\\
\nonumber
\\
	P_{\rm LW} &=& \frac{1}{4\pi^2} \left|\; \int\limits_{C_{\rm BC}} d\tilde{\omega} \;
		\frac{\sin(\tilde{K}z') \, \exp\left[\tilde{\kappa}(z-L)\right]}
		{\tilde{K}\cos(\tilde{K}L) + \tilde{\kappa}\sin(\tilde{K}L)} \;
		\frac{1}{\omega-\tilde{\omega}} \right|^2 \; .
\end{eqnarray}

\begin{figure*}%
        \epsscale{0.5}%
        \plotone{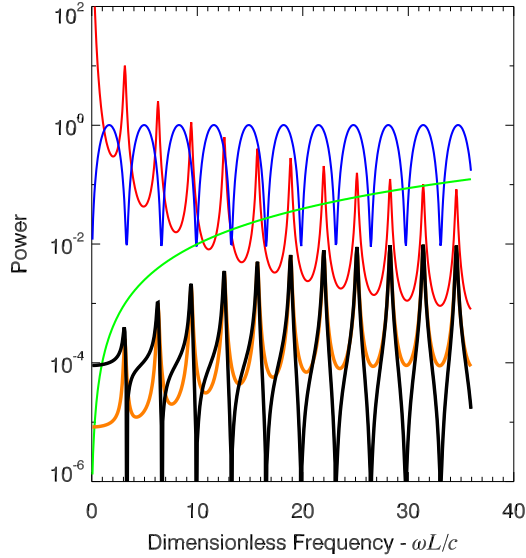}%
        \caption{\small The power spectrum for our simple ``guitar-string" model.
The source height is $z' = 0.95~L$ and the observation height is at $z = 0.99~L$.
The damping time is small compared to the sound-crossing time, $\gamma L/c = 0.1$.
The thick black curve shows the power spectrum of the solution as a function of
frequency, and can be represented in either of two equivalent ways, equation~\eqnref{eqn:Pguitar_cont}
or \eqnref{eqn:Pguitar_mode}. The thin colored curves show the contributions from
three separate factors in equation~\eqnref{eqn:Pguitar_cont}: the effect of the
height of observation (green curve), the reciprocal of the square of the Wronskian
(red curve), and the driving efficiency (blue curve). The total power is the product
of these three factors. The Wronskian embodies the modal structure, and everywhere
there is a resonant mode frequency the Wronskian vanishes, providing a peak with
a Lorentzian profile. The driving efficiency possesses zeros at frequencies lieing
between the modes. Asymmetry arises because the zeros of the driving efficiency do
not fall symmetrically between the mode frequencies. An alternative interpretation
is that the modes represented in equation~\eqnref{eqn:Pguitar_mode} interfere with
each other to produce the asymmetry. The thick orange curve demonstrates that the
line profiles becomes symmetric if one neglects mode interference and assumes that
the total power is the incoherent sum of the power in all modes.%
\label{fig:Power}}%

\end{figure*}

\begin{figure*}%
        \epsscale{1.0}%
        \plotone{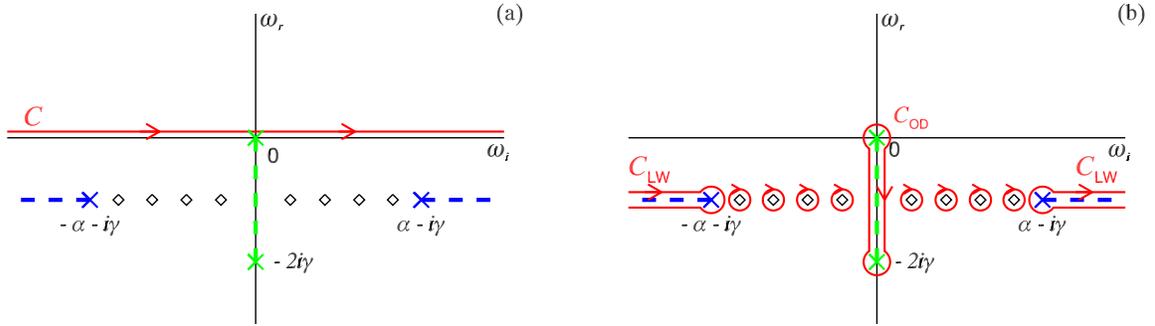}%
        \caption{\small Complex frequency space through which the inverse Fourier
transform integral is performed. (a) The integration contour (red line labeled $C$)
origininally lies along the real frequency axis. (b) For $t>0$, the contour is
deformed downwards and picks up contributions from each of the singularities. The
diamonds mark the location of the poles of the Green's function, which correspond
to the acoustic cavity's complex mode frequencies. The green crosses denote the
branch points that frame the branch cut (green dashed line) associated with overdamped
waves. The integration contour around this branch cut is shown in red and labeled
$C_{\rm OD}$. The branch points and cut associated with leaky waves are indicated
in blue, and the contour around this branch cut is labeled $C_{\rm LW}$.%
\label{fig:omegaSpace}}%

\end{figure*}

\begin{figure*}%
        \epsscale{1.0}%
        \plotone{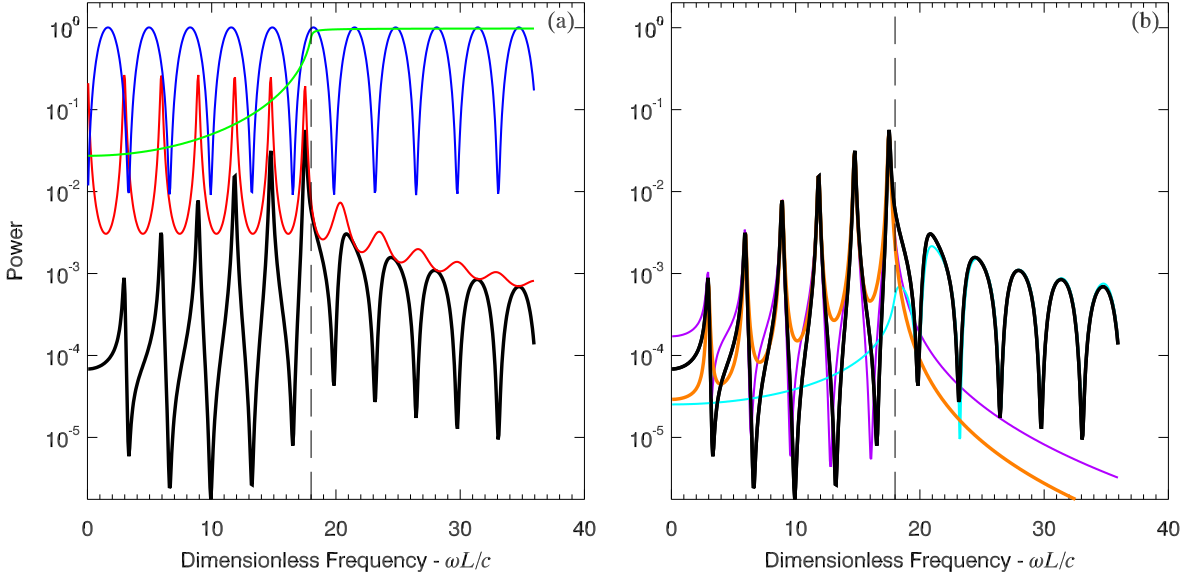}%
        \caption{\small The power spectrum for an acoustic cavity corresponding to
a leaky square-well potential with a cavity depth of $L$. The source height
is $z' = 0.95~L$, the global cut-off frequency is chosen to be $\alpha = 18.0~c/L$,
and the damping parameter is the same as in Figure~\ref{fig:Power}, $\gamma = 0.1~c/L$.
The observation height is at $z = 1.1~L$, just above the upper boundary of the cavity.
(a) The thick black curve shows the power spectrum as a function of frequency obtained
through equation~\eqnref{eqn:PLeaky_cont}. The colored curves have the same meaning
as in Figure~\ref{fig:Power}. The vertical dashed line indicates the global cut-off
frequency $\alpha$. Asymmetry arises because the zeros of the driving efficiency
(blue curve) do not fall symmetrically between the mode frequencies. (b) The power
spectrum (thick black curve) obtained through an eigenfunction expansion,
equation~\eqnref{eqn:PLeaky_modes}. The violet curve shows the power associated with
just the modes, $P_{\rm modes}$, and the aqua curve shows the power due to the continuous
spectrum of leaky waves, $P_{\rm LW}$. The thick orange curve presents the incoherent
sum of the power in the modes, $P_{\rm inc}$.  The asymmetry apparent in the
power is the result of interference between modes and between modes and the leaky
waves.%
\label{fig:LeakyWell}}%

\end{figure*}

\noindent Similar to Figure~\ref{fig:Power}, the thick orange curve shows the incoherent
sum of the power in the modes,

\begin{equation}
	P_{\rm inc} = \sum_{n=-N}^N \left(\frac{K_n\kappa_n^2 c^4}{1+\kappa_n L}\right)^2 \;
		\frac{\sin^2(K_n z') \, \exp\left[-2\kappa_n(z-L)\right]}{\hat{\omega}_n^2 \, \cos^2(K_n L)} \;
		\frac{1-\delta_{n0}}{(\omega-\hat{\omega}_n)^2 + \gamma^2} \; .
\end{equation}




\end{document}